# Structural and Morphological Effect of Ti underlayer on Pt/Co/Pt Magnetic Ultra-Thin Film

M. Turksoy Ocal[1, 2*], B. Sakar[1,2], , I.Oztoprak[1], Z. Balogh-Michels[3], A. Neels[3], O. Ozturk[1, 2]

*(1)* *Physics Dept., Gebze Technical University, Surface Phys. Lab., 41400 Kocaeli, Turkey*
*(2)* *Surface Phys. Lab in Nanotechnology Research Center, Gebze Technical University, Kocaeli, Turkey*
*(3)* *Center of X-Ray Analytics, Empa, Swiss Federal Laboratories for Materials Science and Technology, 8600 Dübendorf, Switzerland*

Pt(xÅ)/Co(5Å)/Pt(10Å) trilayer films were deposited on naturally oxidized Si(111) substrate to investigate Ti underlayer effect on PMA. A small amount of Ti underlayer promoted the magnetic anisotropy to a perpendicular direction for x=8Å and 10Å. Both GI XRD and STM results showed that the Ti underlayer manipulated the growth of trilayer film. The Pt particles preferably grow the neighborhood of the Ti cluster, giving rise to a relatively smooth layer with a fiber texture fcc (111) structure. By contrast, the absence of Ti underlayer leads to clustered, relatively rough, and randomly oriented nano-crystalline growth. Differences in the growth mode, especially roughness, also appeared as intensity loss in XPS spectra due to shadowing effects. Our results indicate that a traceable amount of underlayer can change the magnetic anisotropy of the film by manipulating film growth.

**Key Words:** Perpendicular Magnetic Anisotropy, Multi-Layer Films, Magnetron Sputtering, Ti underlayer, fiber texture growth

# 1. Introduction

Magnetic materials with Perpendicular Magnetic Anisotropy (PMA) have a wide range of applications in magnetic recording, sensing devices, and spintronic. Thus they have been subjected to many types of research published in the literature[1–7]. PMA based thin films owe their popularity to have high thermal stability and the potential to produce smaller and faster devices compared to in-plane magnetic anisotropy-based thin films. Different class of materials such as Ferromagnetic Metal (FM)/Heavy Metal (HM) multilayers (Co/Pt, Co/Pd etc.) [8–10], L10 ordered tetragonal structures (CoPt,FePt) [11,12], HM/FM/HM (Pt/Co/Pt, Ir/Co/Pt etc.) trilayers[13–15] and rare earth-transition metal alloy films(TbFeCo, GdFeCo etc.) [16,17]can be used as a strong PMA structure. Among these materials, the magnetic multilayer films with PMA are great of interest because of their properties such as strong thermal stability, low critical current density, and strong out-of-plane magnetization for nanoscale spintronic application (STT-MRAMs). Tunability of such properties and easily deposition by conventional deposition techniques are another advantage of multilayer PMA films.

In multilayered PMA film (FM/HM multilayers and HM/ FM/HM trilayers), the interface anisotropy responsible for the PMA. Generally, strong spin-orbit coupling at FM and HM interface causes PMA, and PMA strongly depends on FM and HM layer thicknesses [13,18,19], growth parameters [20–23], and structural parameters such as surface roughness[2,24], diffusion[9,25,26], and strain[8,27,28]. Also, it's well known that the textured growth in fcc (111) orientation helps to promote/enhance PMA, and it can be achieved by inserting underlayers/buffer layer such as Pt, Ta, and $TaO_x$ as a template for crystalline growth between the substrate and the bottom layer [8,13,21,23,29,30]. Although the enhancement of PMA by using a buffer/under layer was extensively studied, the reason behind this behavior was not completely emphasized. Some studies attributed this behavior to 2D Frank der Merwe growth with a smoother interface and coherent strains[8,30].

In the present work, the effect of titanium as an alternative underlayer on PMA was investigated in order to produce long life and repeatable PMA thin films In the literature, Ti underlayer was extensively used as an adhesive layer in $Si/SiO_x/Ti/Pt$ type substrate for PZT ferroelectric thin films studies. In these studies, Ti underlayer causes in fcc(111) textured growth acting as a seed layer for PZT thin films[31,32]. Although it is well known that textured growth enhances PMA, to the best of our knowledge, there are only a few studies about Ti buffer/ underlayer layer effect on magnetic anisotropy. In the study of Toyama[33]et al., (Co/Pt)4 multilayer thin films deposited on thermally oxidized Si/SiO$_2$ substrates by E-beam deposition with 3.0 nm thick Ti underlayer on the substrate. Interestingly, the underlayer did not act like only a buffer layer; somehow, it changed the over-layer structure and magnetic properties too. They pointed out that Ti-under-layer has a pinning effect on the some of Co atoms in (Co/Pt)$_4$ multilayers leading to form L$_{12}$ order after annealing. But in their study, multilayer films with and without the Ti underlayer showed in-plane anisotropy. On the other hand, Hyon-Seok Song et al. deposited Ti(dTi nm)/Co(0.3nm)/[Ni(0.7 nm)/Co(0.15 nm)]6/TaN(5nm) multilayers on Si/SiO$_2$(001) substrates by changing titanium layer thickness from 1.5 to 9 nm[34]. They showed that multilayer films have perpendicular magnetic anisotropy for all titanium layer thicknesses. However, Ti thickness is quite higher compared to us, and texture growth effect on PMA was not mentioned.

Herein, Pt/Co/Pt trilayer thin films were deposited with and without Ti (3Å) under layer on naturally oxidized Si(111) substrate by using magnetron sputtering depositions. GI-XRD results showed that even 3Å Ti underlayer could promote Pt(111) textured growth, leading out of plane anisotropy. Morphological and electronic structure differences due to Ti layer were also investigated via UHV-STM and XPS techniques.

## 2. Experimental methods

In this study, two types of film sets were prepared on naturally oxidized(n.o.) Si(111) substrate at room temperature by Magnetron Sputter Deposition Technique. In the first film set, Pt(x)/Co/Pt tri-layer films named pcp sample set were deposited at x=5Å, 8 Å, 10Å Pt layer thicknesses. In the second film set, a 3Å Ti -underlayer was inserted between the first Pt layer and n.o. Si(111) substrate to indicate underlayer effect on magnetic anisotropy, which is named tpcp sample set (Ti/ Pt(x)/Co/Pt). For both film set, Co layer and Pt cap layer thickness were kept at 5Å and 10Å, respectively. Before the deposition process, the substrates were annealed at 450°C for 15 min in the deposition chamber to remove surface contaminations such as hydrocarbons. The base pressure of the deposition chamber was about $1\times10^{-9}$ mbar while working Ar pressure was $10^{-3}$ mbar. RF, DC, and pulsed DC power sources were selected with the sputtering powers of $18.5\times10^{-4}$ W/mm$^2$ (RF), $2.5\times10^{-4}$ W/mm$^2$ (DC), and $12.5\times10^{-4}$ V/mm$^2$ (pulse-DC) for Co, Pt, and Ti depositions, respectively. The distance between substrate and target was kept at 100 mm. The thicknesses and the rates are monitored by a Quartz Crystal Microbalance (QCM) thickness sensor, which is periodically calibrated by the X-Ray Photoelectron Spectroscopy (XPS). The recorded deposition rates were 0.08Å/sec, 0.0533Å/sec and 0.05Å/sec for Co, Pt and Ti, respectively.

Magnetic measurements of the samples were carried out at room temperature by Magneto Optical Kerr Effect (MOKE) system at polar geometry (P-MOKE). To establish correlations between magnetic anisotropy and structural, morphological, and electronic properties in the presence Ti under layer, Pt(10Å)/Co/Pt and Ti/ Pt(10Å)/Co/Pt samples were selected. These samples will be called shortly pcp (without Ti-underlayer) and tpcp (with Ti-underlayer) hereafter. The structural analysis of the the pcp and tpcp samples by X-ray diffraction (XRD) is done by a Bruker D8 Discover Da Vinci diffractometer. Both samples were investigated by two measurement modes: the grazing incidence (GI) out of plane scan and in-plane sector scans. The first type is an established practice, which is especially suitable to investigate thin polycrystalline films. The in- plane sector scans are a set of radial in-plane scans with a certain φ rotation between each individual scans, in this example a 80° wide arc was used with 1° resolution. This method combines the advantages of the in-plane radial and azimuthal scans, that it delivers both the phase and symmetry information. The obvious disadvantage (low intensity) is the requirement of ~100 scans to obtain readily data, as compared to few scans for simpler XRD methods.

The preparation and surface analysis like X-ray Photoelectron Spectroscopy (XPS) and Scanning Tunneling Microcopy (STM) of the thin films were realized in the same UHV cluster system. Elemental analysis and chemical structure of each layer of pcp/tpcp samples were investigated using XPS with Mg Kα-1253 eV as an excitation source. XPS analysis carried out by SPECS system equipped with a PHOIBOS 150 hemispherical analyzer and a double anode x-ray sources under UHV condition (about 1x10-10mbar.) For each the layers in the films, a wide range survey scan with 1eV energy resolution were taken initially in the photoemission spectroscopic characterization. Thereafter, the analysis is focused on the core levels photoemission window scans for a certain element by a high-energy resolution of 0.1eV.

To investigate the morphological surface structure of samples, in situ STM experiments were performed with SPM Aarhus 150 SPECS at room temperature. Each layer of the pcp and tpcp samples was freshly prepared to avoid contamination at the interfaces for STM measurement. All STM images were taken in the constant current mode with a mechanically cut Pt/Ir tip. During STM measurement, bias voltages and tunneling currents were between 0.6-1.25 V and 0.15-0.50nA.

## Results and discussion

**Figure 1** shows the P-MOKE measurements of pcp and tpcp sample sets at room temperature. As shown in the fig.1 (a), all the pcp sample sets exhibited a hard axis hysteresis loop in P-MOKE measurement. By contrast, tpcp samples with x=8Å and 10Å showed a rectangular-shaped easy axis hysteresis loop confirming the perpendicular magnetic anisotropy (PMA) exists except x=5Å (fig.1 (b)). One can conclude that the Ti underlayer between the substrate and first Pt layer promotes magnetic anisotropy to the out-of-plane direction for the certain Pt layer thicknesses. As a nearly perfect rectangular hysteresis loop was obtained in the presence of Ti at 10Å Pt thickness, tpcp (Ti/ Pt(10Å)/Co/Pt) and pcp (Pt(10Å)/Co/Pt) samples were selected to understand underlayer effect.

Since each layer of the magnetic films is ≤10Å, the interface effect between the layers becomes dominant. Thus, the structural and morphological changes at the interfaces are the possible reason behind the PMA in the presence of the Ti layer. Also, it's well known that PMA strongly depends on crystal orientation. Especially, textured growth in (111) direction enhances the PMA[10]. To understand the structural alterations between tpcp and pcp samples, GI-XRD and in-plane-XRD methods were used. The existence of Co layers between relatively thin layers might cause oxidation therefore ex-situ XRD studies carried out on 10Å cap layer to prevent oxidation of magnetic cobalt layer.

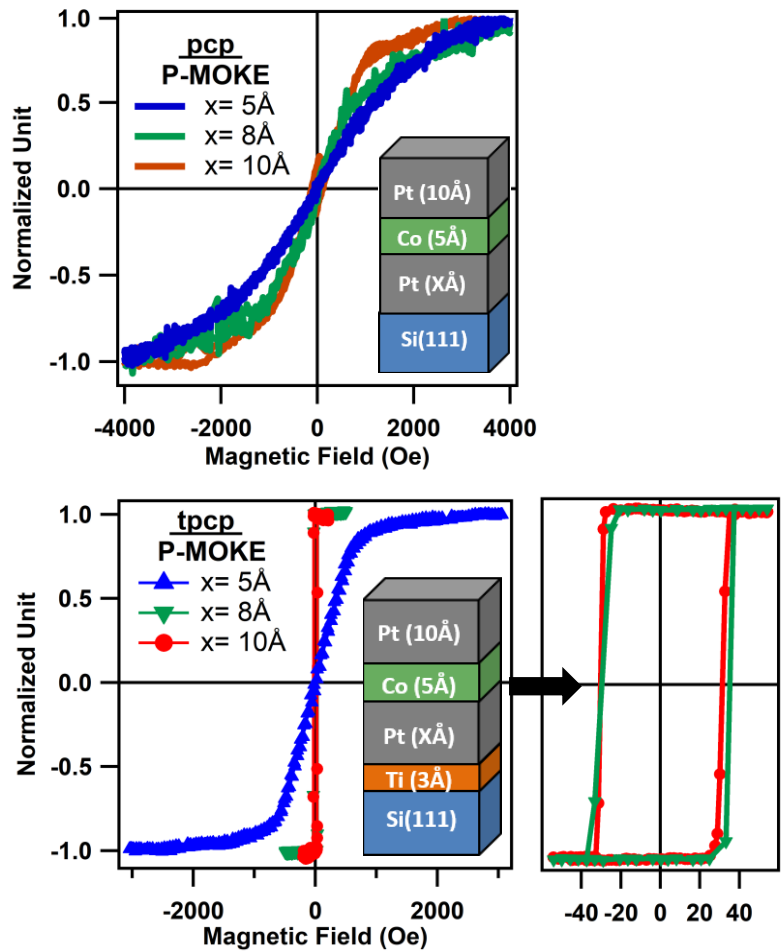

**Figure 1.** Hysteresis Curves measured by MOKE of pcp (top) and tpcp (bottom) samples with different Pt layer thicknesses.

In scan results shown in **Figure 2**, all the significant Pt peaks and some of the Co peaks were present for the pcp sample. However, a clear dominance of the Pt (220) reflection was shown for the tpcp spectrum. Cobalt peaks were not detectable, because either the crystallites are too small or they have an unfavorable orientation for the most intense peaks. For the tpcp sample scans, the Pt (220) (perpendicular to the ($1\bar{1}1$)) is clearly dominant, but instead of separate peaks an almost isotropic ring is visible in the in-plane scan. In **Figure 3**, the separate peaks observed in plane scans are the reflections from a single crystalline substrate while the rings are pointing out for the same reflection from every azimuthal angle.  This is an indication for a fiber-texture, i.e. a (fine-) grained crystalline material, where each crystallite has a $(1\bar{1}1)$ as it's out of plane axis, but it, in plane orientation, is random. The third component of this crystal system $(\bar{2}24)$ is not visible with the XRD instrument. On the contrary, to that, the multiple Pt rings are also observed for the pcp sample, including the (220) rings. These indicate that the pcp sample is a random oriented nano-crystalline film. It was concluded that both these types of scans confirmed the (111) fiber texture for the Pt in the tpcp sample. The diffraction results indicate that the tpcp films are strongly textured, unlike the pcp films, are randomly oriented.

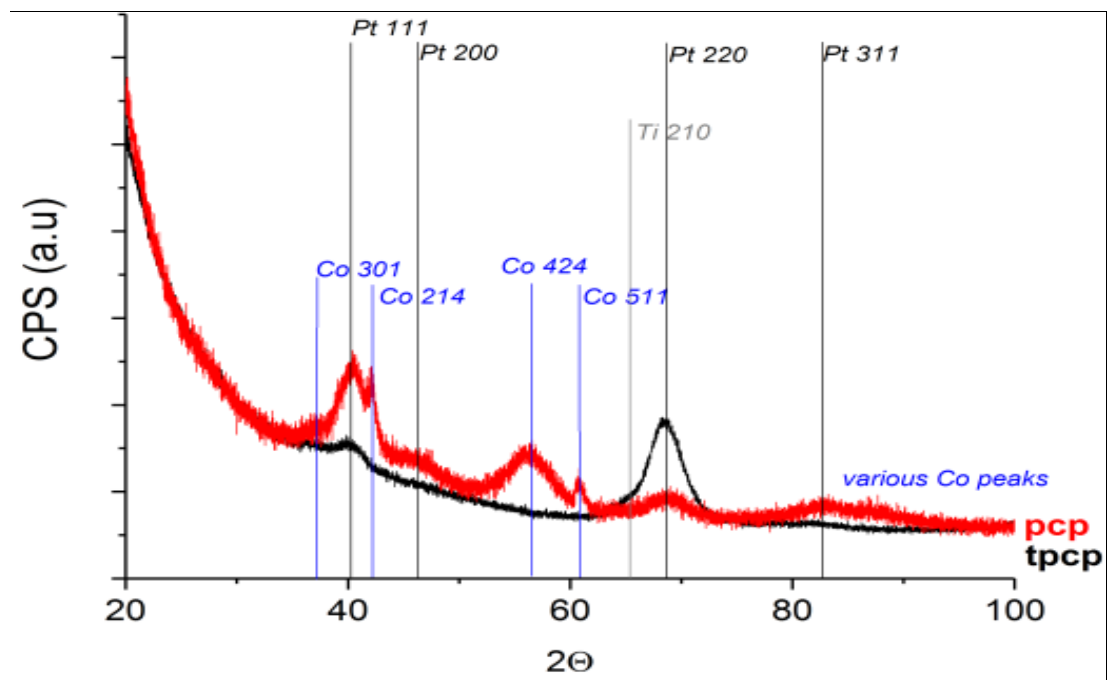

**Figure 2**. Grazing Incidence Angle XRD of tpcp (black) and pcp (red) samples.

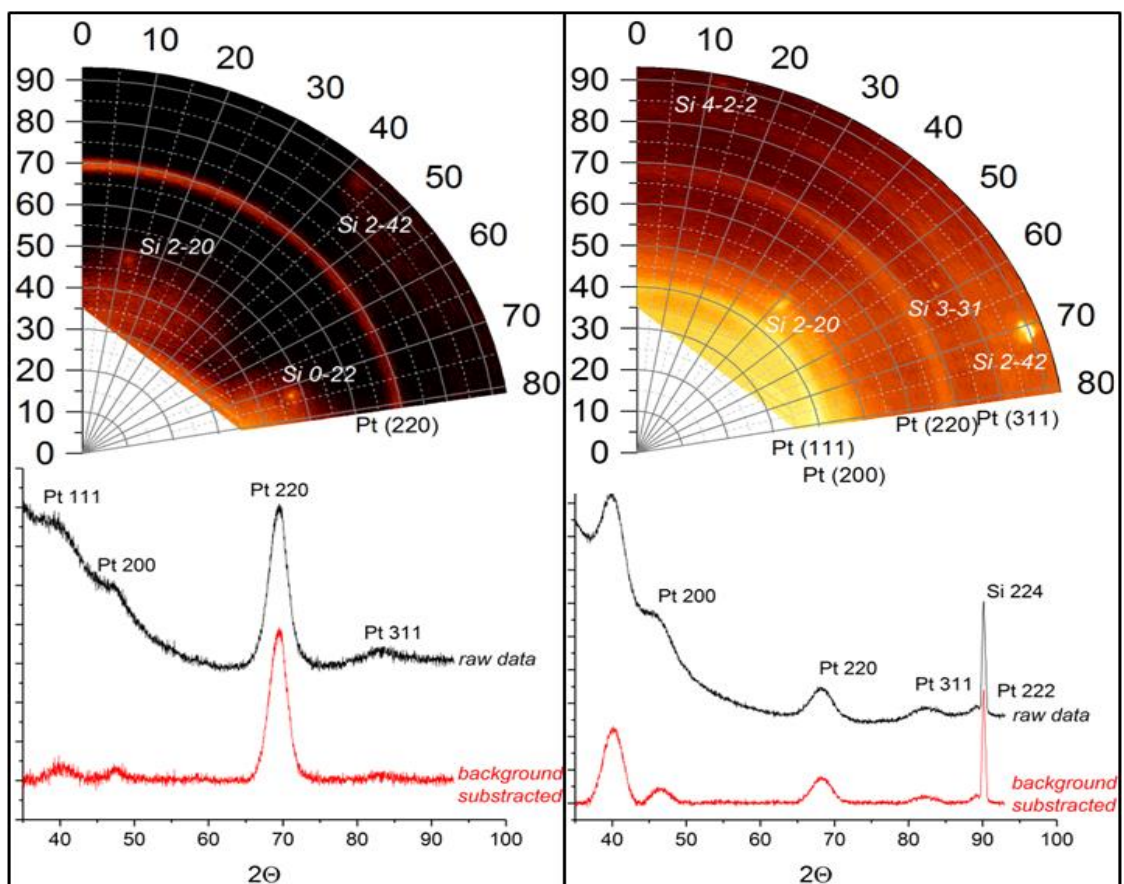

**Figure 3.** In Plane XRD sector scan (top) and line scan (bottom) of tpcp (left) and pcp (right) samples.

Due to texture growth in the fcc (111) orientation, it was not surprising to obtain PMA in the tpcp sample rather than the randomly oriented pcp sample. Similar studies have been carried out using different underlayers such as Ta and TaO$_x$ (3-5nm) [13,30,35]. The underlayer promotes (111) orientation growth and leads to enhance PMA. Fukami et al.[30] explained the contribution of underlayer (Ta / Pt ) to PMA in Co / Ni multilayer film by surface free energy differences of tantalum and platinum. Since the surface energy of tantalum is higher than Pt, they emphasized that the growth of Pt on the Ta layer in Frank der Merwe mode with "thermodynamically preferred" (111) orientation. However, in our case, even the 3Å Ti underlayer promoted the textured fcc (111) growth leading to PMA, although the surface energies of Ti(2.100 J/m2) and Pt(2.475 J/m2) are slightly close to each other [36]. Moreover, the thickness of the titanium is not enough to cover all substrate surface due to the clustering nature of the magnetron sputtering deposition technique without elevated temperature and well-defined substrate. Therefore, it is not the right approach to using the surface energy difference between overlayers in our study.

As mentioned before, for the multilayer thin films, the interfaces between layers become dominantly important. Therefore, to understand the role of Ti under-layer, STM and XPS measurements were taken in-situ without Pt (1nm thick) cap layer after each deposition sequentially. STM images reveal how titanium under-layer changes growth mode and surface morphologies. Figures 5 a and 5b show the STM images for the first and second layers of the pcp samples. It can be seen from Figure 5a that Pt particles formed individually distributed clusters on naturally oxidized Si(111) substrate. RMS (Root Mean Square) roughness value of this layer is about 279.7 pm. After cobalt deposition on 10ÅPt (Figure 5b), the cluster sizes increased due to the pilling up of Co particles over the Pt layer, and the RMS roughness of this layer decreased to 255.7pm. In other words, a relatively smooth surface was obtained after Co deposition. When the 3Å Ti buffer layer was deposited between 10Å Pt film and substrate, the surface morphology changed dramatically. Pt clusters became in contact with each other (Figure 5c) and tended to clothe over the surface instead of clustering. Moreover, some dark spots (as illustrated with black arrows in Fig.5c) started to appear in STM images, comparable to the size of Ti clusters grown on naturally oxidized Si(111). While the dark spot diameters varied in 14Å-22Å, and the Ti cluster radius did around 11Å-18Å (not shown here). Since the amount of deposited Ti is not enough to cover all substrate surface, it is suggested that Pt clusters preferably grew in the neighborhood of Ti clusters interpreted as dark spots.

On the other hand, the RMS roughness value (204.6 pm) decreased by comparing the first Pt layer of the pcp sample. Ti under-layer manipulated the Pt layer growth resulting in a smoother layer. After Co deposition on the first Pt layer of the tpcp sample, surface morphology remained nearly unchanged. As shown in Figure 5d, only dark spots in the STM images became more apparent with the Co deposition. Ti under-layer also manipulated Co layer growth through Pt layer since Pt layer is not entirely buried. RMS roughness (207.8pm) decreased according to the Co layer of the pcp sample but nearly with the same with Pt layer of tpcp sample.

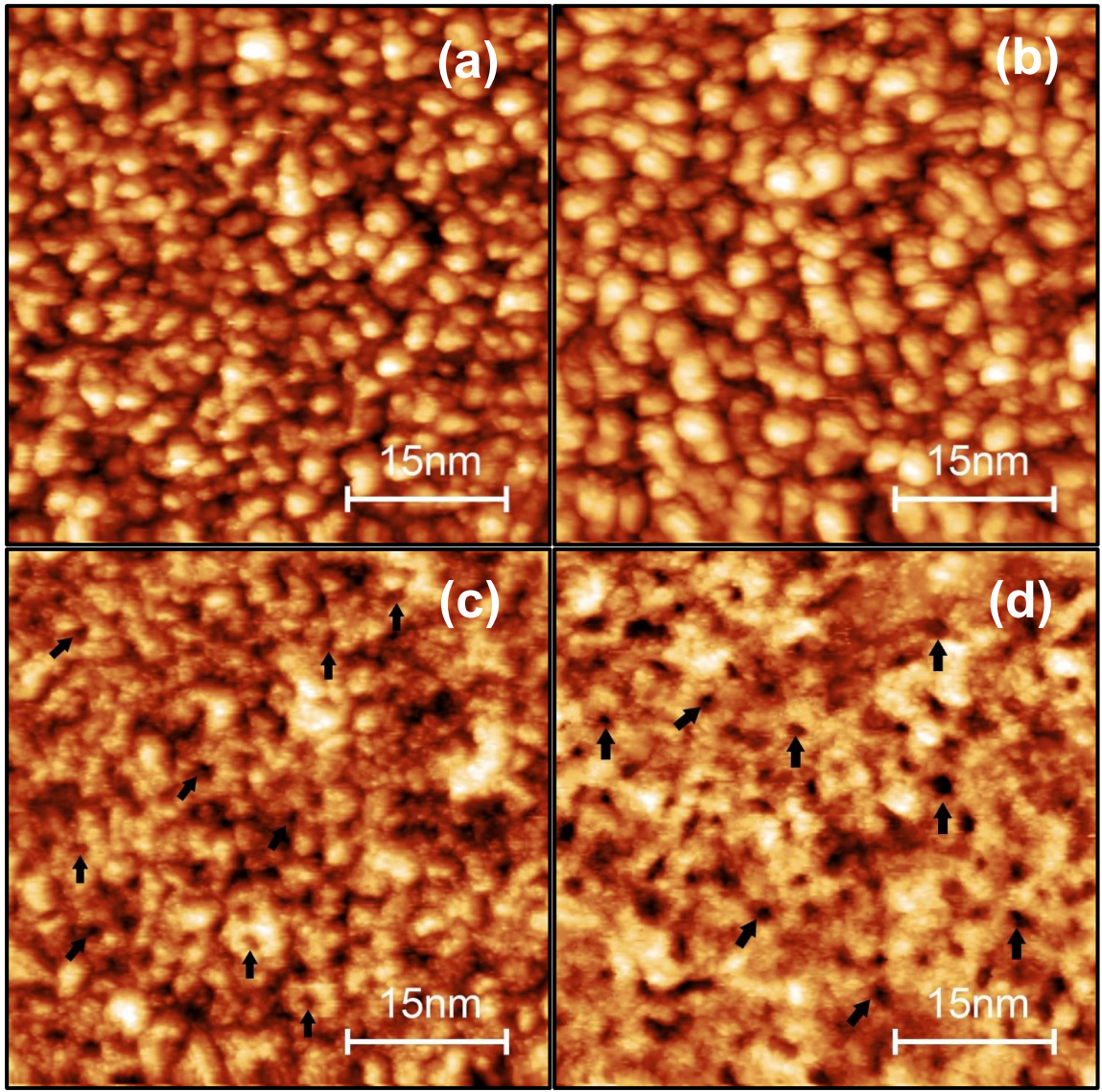

**Figure 5**. Constant current STM images (50nmx50nm) of a) 10ÅPt on no-Si(111) (pcp), b) 5ÅCo/10ÅPt on no-Si(111) (pcp), c) 10ÅPt/3ÅTi on no-Si(111) (tpcp) and d) 5ÅCo/10ÅPt/3ÅTi on no-Si(111) (tpcp)

Both pcp and tpcp samples, Co clusters showed similar morphological order with Pt clusters. In the pcp sample, Co particles accumulated on individually distributed Pt Clusters, which lead to relatively larger Co clusters. On the other hand, the distribution of the Pt clusters in the tpcp sample prevented the accumulation of Co particles, leading to a spread on the Pt layer, and it resulted in relatively small clusters. The growth behavior of the Co layer can also explain the reason for undetected Co clusters at GI-XRD in the tpcp sample.

In addition to fcc (111) textured growth, the smooth interfaces are known to support promoting PMA[37–39]. In our case, the sharper interface (or less rough interface) between Pt and Co layer may also help to promote the PMA in tpcp samples. Besides, Kosuke Suzuki[38] et al. also suggested that a smooth interface enhances the PMA in Co/Pd multilayer film. According to their work, the higher PMA energy can be achieved with a smoother interface even if the lower Pd thickness was used in the Co/Pd multilayer film. They supported to their suggestion by showing the interface roughness effect on PMA via the DV-X a cluster calculation model[37]. They pointed out that PMA decreases due to diffusion at rough interfaces. In accordance with [37,38] studies, it is suggested in our case that the Ti under-layer supports PMA by making both the Pt and Co surface smoother (i.e. sharper interface between Pt and Co layers). On the other hand, M. Bersweiler[39] et al. deposited (Co/Pt) multilayer, with a Pt layer thickness between 1.2 nm–2.2 nm, on Ta/Pt buffer layer over Si substrate. They observed that decreasing the Pt layer thickness leads to an increase of interface roughness, which results in a decrease in PMA. In our case, the absence of PMA for 5Å Pt thickness can be explained also by the increase in interface roughness due to the reduction of Pt thickness.

The presence of the Ti-under-layer does not change only the growth mode of the Pt films, but also it can affect the surface electronically. Therefore, to reveal the role of electronic interactions between particles on interfacial anisotropy, XPS measurements were conducted to study the chemical state of elements at the Co/Pt/Ti interfaces. **Figure 6(a)** shows the Ti 2p high-resolution XPS spectra of the tpcp sample recorded after each layer growth (Pt and Co, respectively). For metallic Ti surface, Ti$2p_{3/2}$ peak should be on 454.1 eV[40], but after the 3Å Ti deposition, the binding energy of Ti$2p_{3/2}$ peak was observed at 454.8 eV which is corresponds to Ti-O compound[41]. Due to high oxygen affinity of titanium, it is interpreted that Ti particles attach to oxygen correlated with the natural oxide of the substrate. After the Pt deposition, the Ti2p core level peak shifted to 0.5eV higher binding energy. The 0.5eV shift on Ti$2p_{3/2}$ was attributed to positive charge left on Ti supported Pt cluster by photoionization process. In our cluster range, this charge can not compensate from the ground due to insulator nature of the substrate surface. The core hole screened by delocalized outer electrons of the cluster, which results in increase of final state energy and decrease of kinetic energy. This kinetic energy loss in final state leads to 0.5eV increase in binding energy. On the other hand, no additional shift occurred on Ti2p peak after the Co deposition indicates no interaction between Ti and Co clusters.

**Figure 7(b)** shows Pt4f high-resolution xps spectra of pcp and tpcp samples, before and after cobalt deposition. The red solid and hallow triangle lines indicate the Pt 4f spectrum of the tpcp sample before and after Co deposition, respectively. Similarly, blue ones correspond to the pcp sample. Although the photoemission peak of $Pt4f_{7/2}$ is on 71.2eV in BE spectra for the metal Pt surface ($Pt^0$) [40], the binding energy of $Pt4f_{7/2}$ peak for the first Pt layer of pcp and tpcp samples were observed at 71.6eV and 71.3eV, respectively. Existence of Ti-underlayer caused a change on the binding energy of $Pt4f_{7/2}$ about 0.3eV. As known, thin films of metallic elements can show a transition from insulator to metallic behavior with the increasing cluster size and metal loading. Moreover, the binding energy approaches to the bulk value with increasing cluster size[42]. In our case, Ti-under-layer increases contact between Pt the particles (see figure 5c) create a more conductive surface in tpcp sample. Therefore, the core hole screening in this layer is similar to the bulk-like Pt surface. However, due to lack of contact between randomly oriented Pt clusters in the pcp sample, the less conductive surface leads to increase of final state energy. It causes more shift rather than tpcp sample.

After Co deposition, the $Pt4f_{7/2}$ peak in tpcp and pcp samples shifted up to 71.5eV and 71.8eV respectively. Both sample indicated 0.2eV shift to the higher binding energy. On the other hand, $Co2p_{3/2}$ peak observed at 778.2eV (as shown in figure c) for both film type. It indicates that Co stays over the Pt Clusters as a metallic form[40], with no electronic interaction to Pt Clusters. The existence of Ti under layer also has no effect on BE of Co2p. However, the intensity differences due to the existence of Ti under-layer were observed in both Pt4f and Co2p spectra. The relative intensities of Pt4f spectra before Co deposition in **Figure 6(b)** are different in both tpcp and pcp. The spectrum of the 10 Å Pt grown directly on the substrate (pcp) has 24% lower intensity than the same thickness of Pt grown on the 3 Å Ti.

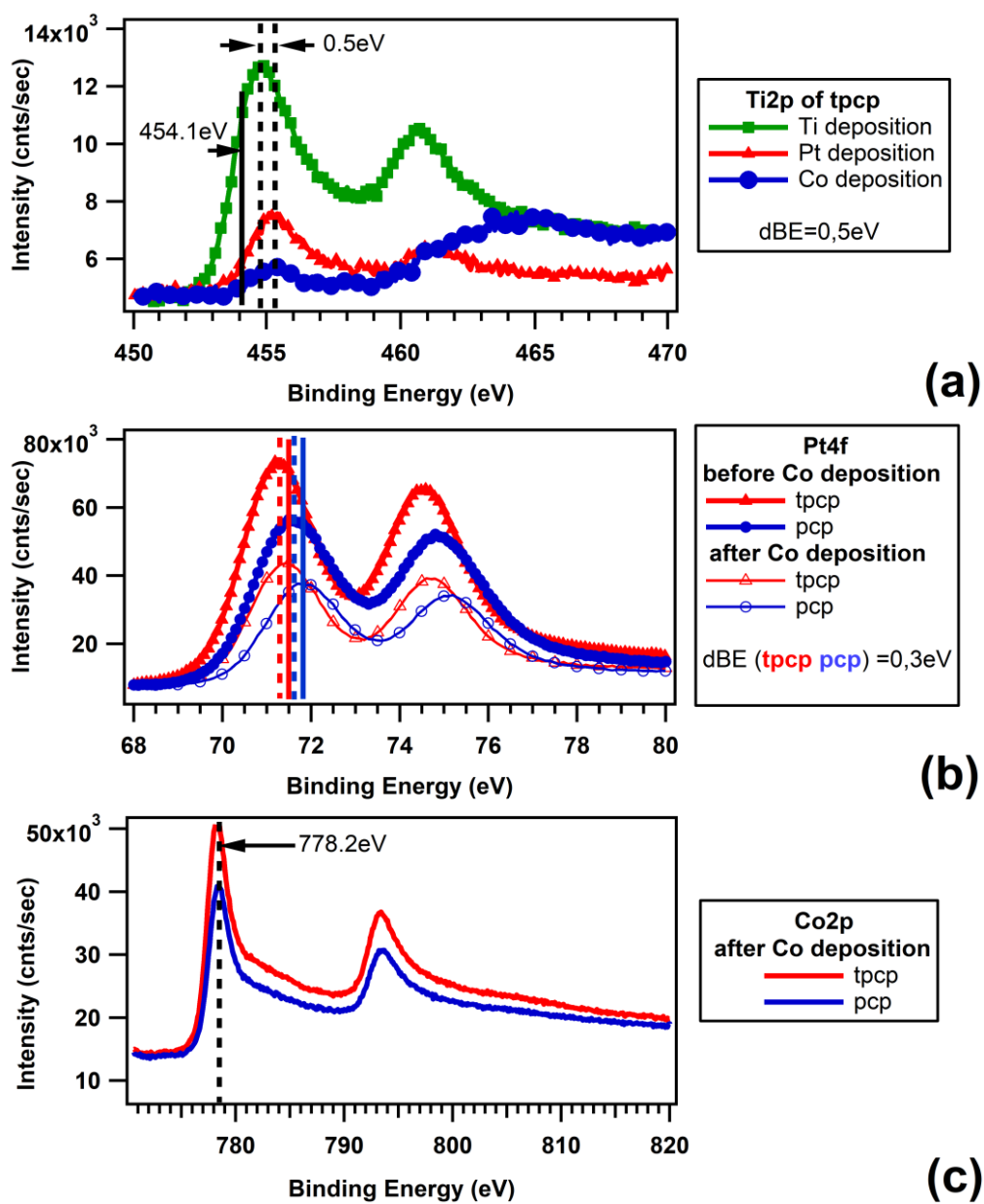

**Figure 7.** High resolution Ti2p (a), Pt4f (b) and Co2p (c) xps spectrum of tpcp and pcp samples.

As well known, the intensity of the photoemission spectroscopy is very sensitive to the surface roughness[43]. The more the surface roughness causes the less the photoemission intensity due to shadowing effect. We suggest that the decreasing spectral intensities in Pt4f and Co2p high resolution spectra indicate a change in the growth modes of Pt and Co layers which were also confirmed from both XRD and STM results.

Table 1. List of binding energies of each elements.

| Layer | Sample | Ti 2p (BE) | Pt 4f (BE) | Co 2p (BE) |
|---|---|---|---|---|
| Ti (3Å) | Tpcp | 454.8eV | - | - |
| Pt (10Å) | tpcp | 455.3eV | 71.3eV | - |
|  | pcp | - | 71.6eV | - |
| Co (5Å) | tpcp | 455.3eV | 71.5eV | 778.2eV |
|  | pcp | - | 71.8eV | 778.2eV |

In summary, our experimental results point out that the dominant reason behind PMA observed in tpcp samples is morphological and structural effects rather than electronic interactions at the interface. It is suggested that the improved interface roughness, in other words, a sharp interface between the Pt and Co layer, leads to promote PMA.

## 4. Conclusions

We investigated the Ti-under-layer effect on the magnetic properties of the Pt/Co/Pt trilayer film prepared by Magnetron Sputtering Deposition onto the natural oxide Si(111) single crystal substrate. It was observed that a small amount of Ti(3Å) underlayer could promote the magnetic anisotropy out of the plane direction. Ti underlayer provokes the growth of the layers leading to a relatively less rough layer surface. The change in the magnetic anisotropy in the presence Ti underlayer is attributed to smoother layer growth in texture fcc (111) direction. Moreover, critical platinum underlayer thickness (more than 5Å) is also important to obtain a well-defined smooth surface leading to PMA.

## Acknowledgments

The work at the Gebze Technical University-Surface Physics Laboratory was supported by several projects; "European Metrology Program coordinates one of projects for Innovation and Research (EMPIR-15SIB06 NanoMag)" and "the Scientific and Technological Research Council of Turkey (TUBITAK- 119M287)". GI-XRD works were completed at Empa, Swiss

Federal Laboratories for Materials Science and Technology.

metal clusters: The d bands of silver and palladium", Physical Review B, 33 (8), 5384–5390.

[43] Martín-Concepción A.I., Yubero F., Espinós J.P., Tougaard S., (2004), "Surface roughness and island formation effects in ARXPS quantification", Surface and Interface Analysis, 36 (8), 788–792.

19